# Intrinsic ferromagnetic impurity phases in SmFeAsO$_{1-x}$F$_x$ detected by μSR

S. Sanna,[1] R. De Renzi [1], G. Lamura[2], C. Ferdeghini[2], A. Martinelli[2], A. Palenzona[2], M. Putti[2], M. Tropeano[2] and T. Shiroka[3]

___________________________________________________________________________

We report about μSR measurements on SmFeAsO$_{1-x}$F$_x$ which helped us to identify the signature of diluted ferromagnetic inclusions, ubiquitous in the iron pnictides. These impurities are characterized by a Curie temperature close to room temperature and they seem responsible for a non negligible magnetic relaxation of the implanted muons, that should not be confused with intrinsic pnictide properties.

___________________________________________________________________________

Author for correspondence: roberto.derenzi@unipr.it - Tel: +390521905279

__________

[1] CNISM and Physics Department, University of Parma, Parma, Italy

[2] CNR-INFM-LAMIA and University of Genova, Genova, Italy

[3] Laboratory for Muon-Spin Spectroscopy, Paul Scherrer Institut, Villigen PSI, Switzerland

# 1. INTRODUCTION

The discovery of high temperature superconductivity in doped iron pnictides [1] brought widespread attention also to their magnetic properties. Often these materials contain other magnetic components, besides the FeAs layers of the pnictide, such as for instance a magnetic rare earth (RE). This is specifically the case for the materials presently holding the record of the highest critical temperatures, $SmFeAsO_{1-x}F_x$ [2] and $NdFeAsO_{1-x}F_x$ [3]. The interplay of the RE magnetism with that of Fe ions and with the superconductivity is of course of primary interest.

A further common, extrinsic source of magnetism is due to small fractions of impurity phases containing Fe. Small deviations from the ideal stoichiometry result for instance in traces of different $Fe_xAs_y$ phases, difficult to identify when they are below the detection threshold for powder x-ray diffraction. These magnetic inclusions dominate the macroscopic magnetization and they are also visible in $^{75}As$ NMR [4].

In this paper we present clear evidence of the presence of ferromagnetic inclusions, in the form of diluted clusters within a matrix of Sm-based pnictide. We could identify their specific effect on the µSR relaxation function and separate it from the intrinsic magnetic and superconducting properties of these new compounds.

# 2. EXPERIMENTAL PART

Bulk polycrystalline pellets of $SmFeAsO_{1-x}F_x$ were prepared by a two-step standard synthesis, as reported elsewhere [5]. During the preparation the samples were sealed in a tantalum capsule which prevents the partial reaction of fluorine with the sealed quartz vessel, ensuring that the F content scales with the nominal one, which is intended henceforth both as an upper limit to the real content and as a sample label. We produced a series of samples which differ in fluorine content $x$, ranging from zero (for the undoped parent compound) up to 0.20. The same samples are already the subject of another publication [6]. In Zero Field (ZF) µSR experiments the implanted muons probe the whole sample volume, detecting both the static and the fluctuating internal fields. In

Longitudinal Field (LF) μSR experiments an external magnetic field is applied along the initial muon spin direction to distinguish the two internal fields. Both types of experiments were performed on the GPS spectrometer of the SμS laboratory (PSI, Switzerland).

## 3. RESULTS AND DISCUSSION

The intrinsic magnetic order parameter of the $SmFeAsO_{1-x}F_x$ compounds has two components, involving separate ordering of the Fe and Sm sublattices. The magnetic transition temperatures for these samples, $T_m$ and $T_{Sm}$ respectively, are reported in Tab.I, together with the superconducting critical temperatures, $T_c$. The main transition temperature, $T_m$, is determined as the temperature below which a precession of the muon starts to appear in zero applied field experiments (ZF-μSR) [6]. The onset of a magnetic order in the Sm sublattice is demonstrated in the $x = 0$ compound by the identification of satellite precessions, corresponding to the vector addition of a further internal field. The three samples with the lower doping ($x = 0, 0.05, 0.075$) display Fe magnetic ordering with rather sharp transitions and no experimental sign of bulk superconductivity. Below $T_{Sm}$ a broad distribution of additional internal fields signal the presence of static Sm magnetic moments. At very low temperatures either static or slow fluctuating Sm magnetic moments are present also in the $x \geq 0.10$ compositions, where they give rise to a strong increase of the muon relaxation rates. We have shown elsewhere [6] that the results summarized in Tab. I demonstrate the coexistence of superconductivity and *static* magnetism only for a very narrow range of compositions around $x=0.085$.

We now concentrate on a temperature region above all the previously mentioned transitions, where both Fe and Sm magnetic moments are fluctuating on a short time-scale, and hence they cannot be expected to contribute to a *static* internal field. Figure 1 shows the muon asymmetry, $A(t)$, of two representative samples ($x = 0.085$ and $x = 0.20$), for different temperatures and externally applied fields. Let us first focus on the zero field data, that may be fitted by a simple two component model, $A(t) = A_1 \exp(-\lambda_1 t) + A_2 \exp(-\lambda_2 t)$. Figure 2 shows the zero field (ZF) muon relaxation rates $\lambda_1$ as a function of temperature for all our samples. The four arrows indicate the two samples and two temperatures where the data sets of Fig. 1 were collected.

The measured relaxation rates are in the range $0.5 < \lambda_1 < 1.5$ μs$^{-1}$ and $0.03 < \lambda_2 < 0.09$ μs$^{-1}$. The values of the component with the larger asymmetry, $\lambda_1$, are somehow surprising. They are much smaller than those characteristic of static Fe and Sm order, and, close to the magnetic transition they could be justified by slow electron spin fluctuations. However spin fluctuations eventually become very fast with increasing temperature and this relaxation mechanism should disappear, whereas the experimental rates $\lambda_1$ remain much larger than the values expected from the pure nuclear moments (dashed line in Fig. 2). Relaxations of the same order of magnitude are seen in most of the μSR reports [7-9], although they are never explicitly discussed.

Furthermore, these rates display hysteresis, i.e. they increase if an external field is applied and then turned off again. Open symbols in Fig.2 show a typical increase in relaxation rate due to the sample magnetic and thermal history (exposure to 0.6T). Notice that, all the other data points follow a a smooth moderate temperature dependence at higher $T$, inconsistent with spin fluctuations.

The static or fluctuating nature of the muon spin relaxation may be determined experimentally. These two cases are distinguished by LF-μSR, by applying a magnetic field along the initial muon spin direction. While in the static case a field of the order of $\lambda_1/2\pi\gamma$ ($\gamma = 135.5$ MHz/T, muon giromagnetic ratio) is sufficient to quench the relaxation, which displays a characteristic Kubo-Toyabe shape [10], in case of dynamic fluctuations the applied field has only a minor effect on the relaxation, which remains exponential.

The results of these measurements are shown in Fig. 1 along with the ZF data. The lower left panel ($x = 0.20$, $T = 20$ K) is attributed to a dynamic relaxation (i.e. fluctuating moments), since a double exponential function (solid lines) represents always the best fit, irrespective of the applied field. These data are measured at $T = 20$ K, where conspicuous Sm fluctuations are still expected [8]. However, all the other three panels are nicely fitted by Lorentzian Kubo-Toyabe functions, multiplied by a modest independent exponential relaxation. We recall that a Lorentzian shape is appropriate for a distribution of *diluted* static moments, such as those of small ferromagnetic clusters. The muon relaxation rates show a strong reduction already at $\mu_o H = 10$ mT, which proves that they arise from a distribution of static moments, reflecting internal fields of the same order (corresponding to the dipolar field of a distant magnetized cluster). Their persistence up to a rather high temperature and the similar behaviour for quite different compositions (from the parent antiferromagnet to the optimal superconductor) indicates their extrinsic origin connected with a magnetic impurity phase characterized by a high Curie temperature. The hysteretic behaviour clearly supports this interpretation.

The presence of diluted impurities throughout the sample volume is detected also by SEM analysis. Fig. 3 shows a back-scattered SEM image (with analysis) for the x = 0.10 sample, selected as representative of the whole class. Polygonal crystals of $SmFeAsO_{0.90}F_{0.10}$ represent the vast majority of the bulk, surrounded by thin layers of $Fe_2As$ (dark grey phase), as checked by Energy Dispersive X-ray Spectroscopy (EDS). Aggregated particles of SmOF (light grey phase), randomly distributed throughout the whole sample volume, are also detected by a combination of EDS and x-ray diffraction patterns; black regions are pores embedded by resin. Neither of these identified phases is ferromagnetic so that the origin of the observed high temperature relaxations remains unassigned, most likely due to spurious Fe clusters.

We notice a final intriguing aspect, which emerges from Fig. 2: whereas the high temperature behaviour is similar for all compounds, the magnitude of the relaxation rates displays a marked dependence on sample composition. Rates are small for both the parent compound and the optimally doped superconductor, but they peak for the compositions corresponding to or close to the magnetic-superconducting transition. This correlation indicates a magnetic coupling between the bulk matrix and the static dipolar field of the inclusions. Such a coupling could for instance be indirectly connected with the observed nanoscopic coexistence of antiferromagnetism (AF) and superconductivity [6]. This peculiar coexistence, observed also in cuprates [11-13] implies large boundaries between nanoscopic size domains of the two intrinsic phases. The small finite size of the AF domains produces in turn a spin imbalance, hence a net moment which may easily couple with the dipolar magnetic fields.

## 4. CONCLUSIONS

We provided evidence of the presence of diluted ferromagnetic inclusions in samples of

SmFeAsO$_{1-x}$F$_x$ in agreement with previous NMR observations of the same compounds [4]. Muon spin relaxation measurements show that these impurities are responsible of a non negligible magnetic contribution up to room temperature, in addition to the intrinsic magnetic properties of the samples. Our data suggest that such spurious effects should be carefully considered, since at the state of the art they seem to regard many iron pnictides. A subtle magnetic coupling with the intrinsic properties of the pnictide may be justified in terms of the observed nanoscopic coexistence of magnetic and superconducting phases.

Table I

Transition temperatures of SmFeAsO$_{1-x}$F$_x$ samples as a function of F doping. $T_m$ and $T_{Sm}$ were determined from zero field µSR, while $T_c$ from SQUID magnetometry [6]

| x | $T_m$ (K) | $T_{Sm}$ (K) | $T_c$ (K) |
|---|---|---|---|
| 0 | 139.0(5) | 5.0(5) | - |
| 0.05 | 96(1) | 5(1) | - |
| 0.075 | 82(2) | 5(1) | - |
| 0.085 | 22(2) | 5(1) | 14(2) |
| 0.10 | - | 5[a] | 41(1) |
| 0.15 | - | 5[a] | 44(1) |
| 0.20 | - | 5[a] | 50(1) |

[a] Corresponding to the onset of an increased low-temperature muon spin relaxation

**FIGURE CAPTIONS**

Figure 1. (Color online) ZF and LF muon asymmetries in SmFeAsO$_{1-x}$F$_x$ for $x = 0.085$ and $x = 0.20$ at two different temperatures. Solid curves are best fits (see text), the dashed line indicates the expected magnitude of the nuclear dipolar contribution.

Figure 2. (Color online) Temperature dependence of the ZF relaxation rates $\lambda_1$ in SmFeAsO$_{1-x}$F$_x$ for different doping levels: $x = 0$ (◄), 0.05 (■), 0.075 (♦), 0.085 (●), 0.10 (▲), 0.15 (▼), 0.20 (►). The open symbols correspond to measurements after an external field was applied and then turned off again. The lines are guides to the eye, for $T \gg T_m$, $T_{Sm}$ and no magneto-thermic history. The arrows represent the temperatures displayed in Fig.1.

Figure 3. Typical SEM image (SmFeAsO$_{1-x}$F$_x$, $x = 0.10$) showing the presence of SmOF clusters randomly distributed throughout the sample volume.

Figure 1

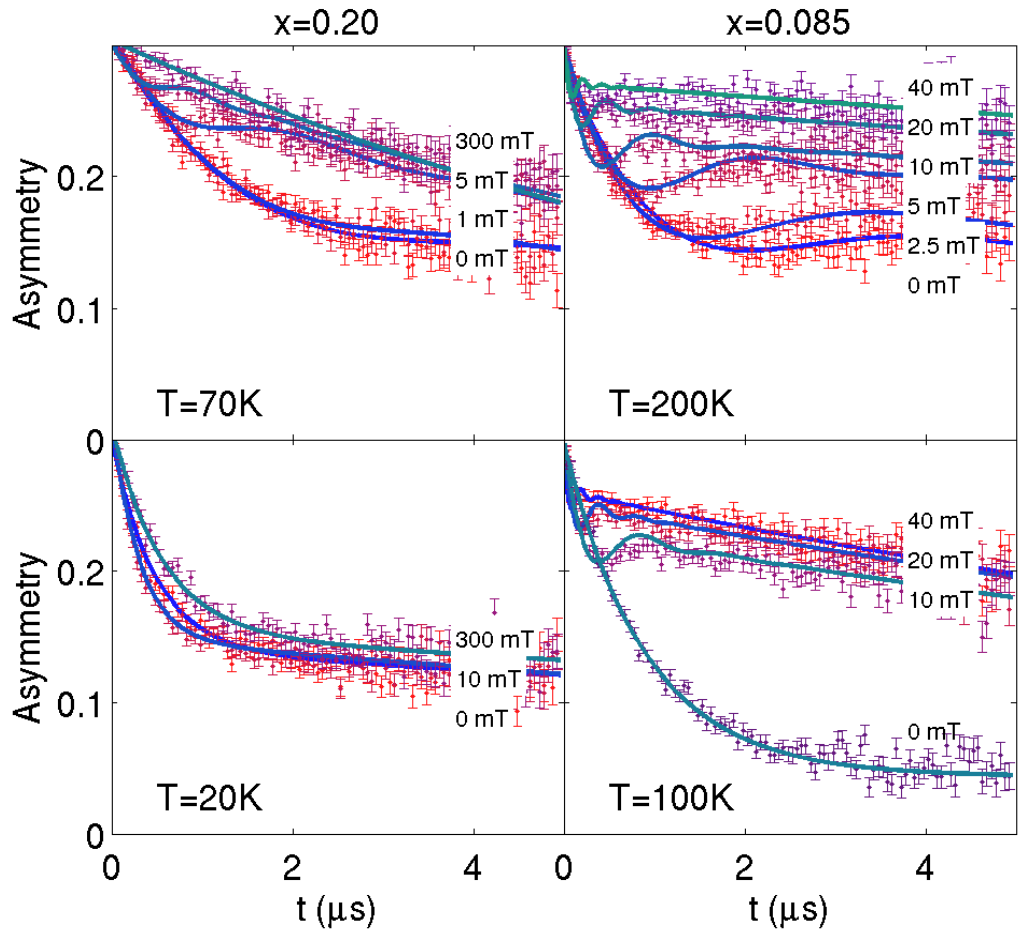

Figure2

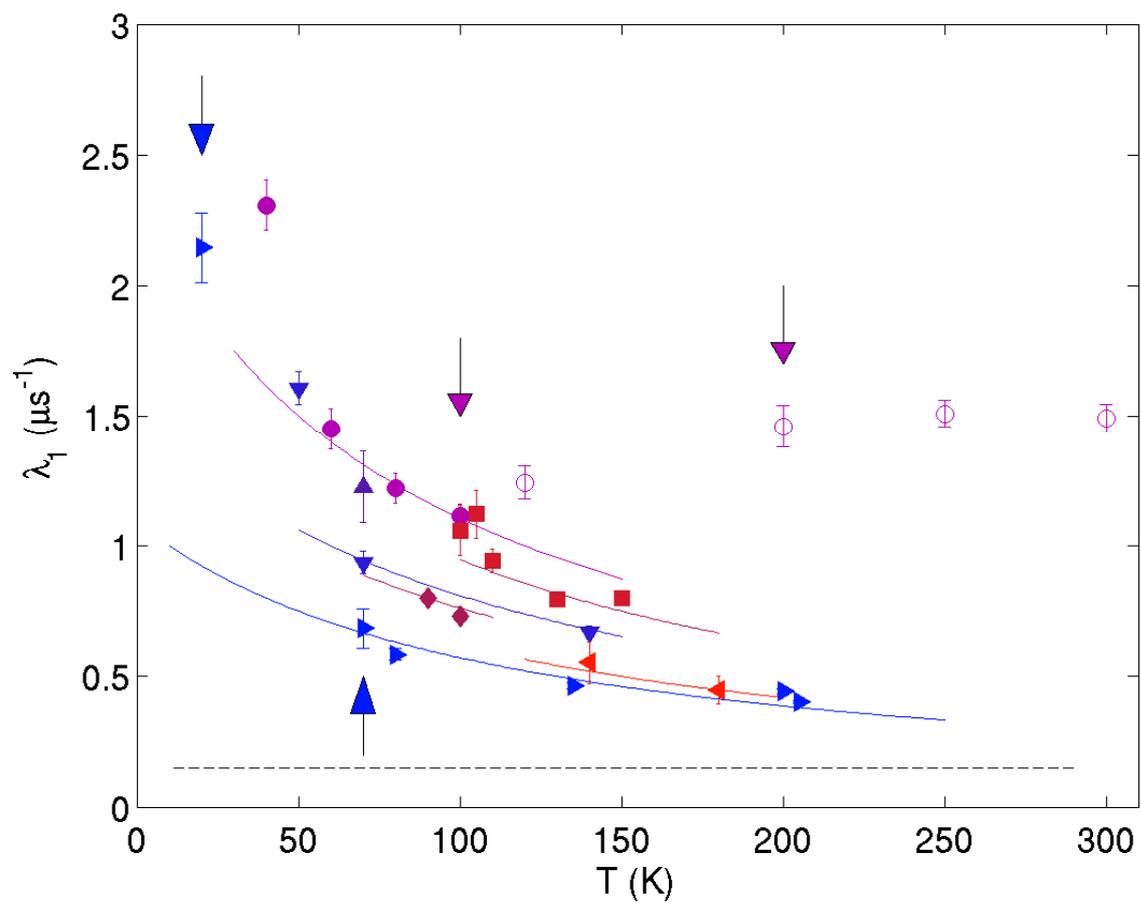

Figure3

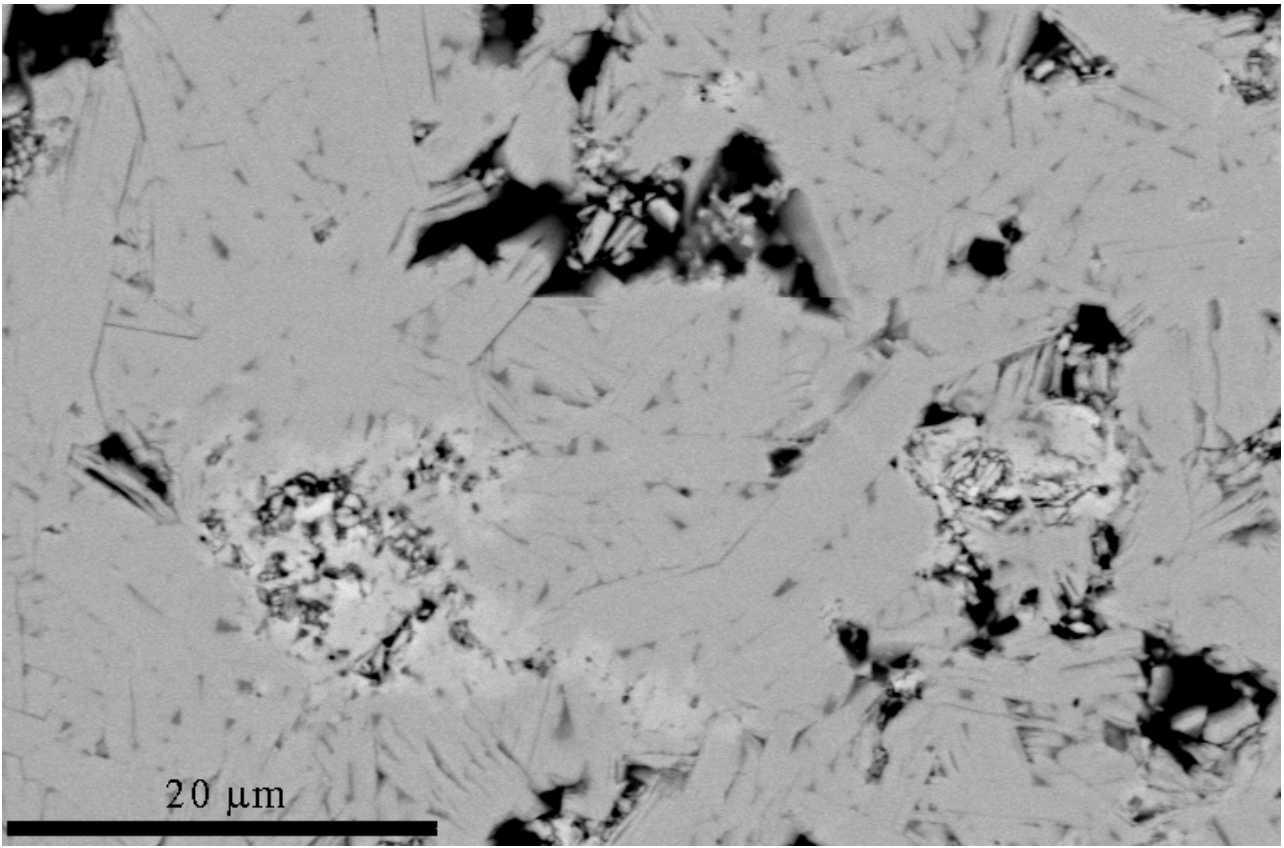